**Longitudinal bunch profile diagnostics in the 50fs range using coherent Smith-Purcell radiation.**

Nicolas Delerue, George Doucas, Ewen Maclean and
Armin Reichold
*The John Adams Institute, Dept. of Physics, Denys Wilkinson Building, University of Oxford, Oxford, UK.*

## Abstract

We have considered the possibility of using coherent Smith-Purcell radiation for the single-shot determination of the longitudinal profile of 50 fs (FWHM) long electron bunches. This length is typical for the bunches currently produced by Laser Wakefield Acceleration and is at the limit of what is achievable by alternative techniques, such as Electro-Optic sampling. It is concluded that there are no obstacles, either theoretical or experimental, in the implementation of this technique. A set of three gratings, with periods of 15, 85 and 500 $\mu$m, will produce detectable energy in the wavelength region 10-1000$\mu$m, which should be adequate for the reconstruction of the bunch shape by the Kramers-Kronig technique. For bunch charges of $10^9$ electrons, or more, the radiated energy can be detected by room temperature pyroelectric detectors. The limits of possible extension of the technique to even shorter lengths are also considered.

## I. INTRODUCTION

Recent advances in Laser Wakefield Acceleration (LWA) [1,2] have brought closer the realisation of compact electron accelerators, with electric field gradients (10-100 GV/m) that are orders of magnitude higher than those achievable with conventional acceleration techniques. One of the most interesting characteristics of the electron bunches produced by LWA is their very short bunch length which is lies in the few femtosecond range.

The repetition rate of the lasers driving these beams is currently low and, moreover, there are significant variations from bunch to bunch. This, in turn, necessitates the development of novel electron beam diagnostic techniques that are capable of measuring the main beam parameters (e.g. emittance and transverse bunch profile) in a single shot. The measurement of the bunch profile in the third dimension (time or 'longitudinal') is also very important, especially if the laser-produced bunch is going to be used as a 'driver' for a Free Electron Laser (FEL). The determination of the time profile of fs-long bunches can, in principle, be achieved by one of the following techniques: (a) the use of a transverse deflecting cavity, where the bunch is deflected out of the beam path and its longitudinal profile is 'flipped' into a transverse dimension [3-5]. (b) Electro-optic (EO) sampling of the bunch [6, 7]. (c) The use of a coherent radiative process, whereby the bunch is made to emit a small amount of electromagnetic radiation and its time profile is then reconstructed from the measurement of the spectral distribution of this radiation. The first of these techniques is applicable down to the few fs range but is destructive and cumbersome. EO sampling has been pursued by a number of groups and has been demonstrated down to about 65 fs, this being the limit imposed by the material properties of the currently available EO crystals [7]. The radiative processes such as transition, diffraction and Smith-Purcell (SP) radiation [8], are relatively simple to implement but they are not 'direct' processes in the sense that the time profile of the bunch has to be reconstructed from the measured spectral distribution of the radiation. To the authors' knowledge, no radiative process has been used for bunch lengths of the order of a few tens of femtoseconds.

In the present paper we consider the possibility of using coherent SP radiation for the *single-shot* determination of the time profile of electron bunches that are 50 fs long. This figure has been chosen because it is within the range of currently produced LWPA beams and is also just below the limit of what is achievable with the EO technique. The paper is divided into two parts: the first is a theoretical study of what can be expected from coherent SP radiation in terms of grating selection, output energy and, also, capability of discrimination between different bunch profiles. We note, incidentally, that although the terms 'bunch length' and 'bunch profile' are used interchangeably, there is a difference between the two and the determination of the (time) profile is obviously the more difficult task. The second part of the paper is devoted to experimental issues such as grating manufacture, availability of suitable optical filters and the choice of the appropriate detectors. The paper is not intended to be a 'design study' for a specific device but rather as an investigation of the opportunities offered by SP radiation as a diagnostic tool in the fs region.

## II. OVERVIEW OF SMITH-PURCELL RADIATION.

### 1. Determination of the time profile.

The details of Smith-Purcell radiation, both theoretical and experimental, have been described in a number of publications. A partial list of publications during the last 20 years can be found in [9-19]. This section only covers the features that are essential for the development of the arguments presented in the sections that follow.

The term is used to describe the radiation emitted from the surface of a periodic metallic structure (a grating) when a charged particle beam is travelling past the grating. Although any charged particle beam will give rise to this radiation, it is assumed that the particles are electrons and that they are highly relativistic ($\beta \cong 1$). A schematic of the process and the axis convention used in this paper are shown in Fig.1. The beam is travelling in the *z*-direction, while the grating grooves are in the *y*-direction. The grating acts as a dispersive element and the wavelength of the radiation will depend on the angle of observation ($\theta$), relative to the beam direction. In most experimental situations the azimuthal angle ($\phi$) is close to the vertical, in which case the relationship between wavelength ($\lambda$), grating period ($l$) and $\theta$ is given by the well-known formula:

$$\lambda = \frac{l}{n}\left(\frac{1}{\beta} - \cos\theta\right) \qquad (1)$$

where $n$ is the order of the radiation. It is assumed throughout this paper that $n$=1, although it should be noted that measurable energy may also be available in higher orders. The range of wavelengths is determined by the choice of grating period and is, thus, a parameter that can be chosen by the experimenter.

The relative magnitudes of the radiation wavelength and the bunch length determine the intensity of the radiation. If the wavelength is shorter than the bunch length, the radiation is incoherent and is proportional to the number of the electrons in the bunch. If the wavelength is approximately equal to or greater than the bunch length, then the radiation is coherent. In this case, the radiated energy is enhanced significantly and becomes proportional to the square of the number of electrons in the bunch. Therefore, it is possible to shift the SP radiation into the coherent regime through a suitable choice of the grating period. Assuming that the spectral distribution of the coherent SP radiation has been measured, it is then possible to reconstruct the time profile of the bunch that gave rise to this distribution. This can be done either by comparison of the distribution with 'templates' that have been calculated for a number of known, simple bunch profiles, or by applying the Kramers-Kronig (KK) analysis technique [20-23] for retrieval of the minimum phase of the bunch. The latter technique is preferable but, for this reconstruction to be meaningful, it is essential to carry out measurements over the widest possible range of wavelengths. A detailed description of the application of the KK analysis to the problem of bunch shape reconstruction and of its limitations can be found in [24]. Moreover, the requirement for 'single-shot' capability means that these wavelengths must be measured simultaneously. Therefore, it is implicitly assumed that there is going to be an array of detectors capable of detecting SP radiation over a broad angular (i.e. wavelength) range. Even so, it will always be necessary to interpolate between the measured data points and extrapolate beyond. SP radiation is particularly advantageous in this respect because it is possible to achieve this wide coverage of wavelengths through the simultaneous use of more than one grating, each having a different periodicity. It is worth noting, incidentally, that a *rough* estimate of the bunch length only (but not of the profile) can be obtained from measurements with a single grating, by observing the approximate wavelength at which the radiation starts rising steeply from the incoherent level towards the fully coherent regime. We have based our study on the assumption that the experimental arrangement can accommodate three gratings.

## 2. Theoretical background.

The application of SP radiation to the problem of bunch profile reconstruction requires an appropriate theoretical description of the emission process, in order to allow for estimates of the expected radiated energy at various wavelengths. Although the problem has attracted considerable attention over the last few decades, there is as yet no universally accepted theory that describes the emission process. This is due, to a significant extent, to the fact that most of the available data are in the far infrared part of the spectrum where precision measurements are difficult. The current situation can be summarised as follows.

SP radiation can be thought of as arising from the diffraction by the grating of the evanescent waves associated with the travelling electron. The origin of this idea goes back to the early 60's [25] and has been refined over the subsequent years [26-31]. Its most recent form uses the electric field integral equation (EFIE) method to calculate the energy radiated from a grating of finite length. Good agreement between theory and measurements has been observed in experiments carried out using a 15 MeV electron beam [18]. Alternatively, SP radiation can be considered as originating from the acceleration of a distribution of charges, the distribution in question being that induced on the grating surface by the travelling electron [32-34]. This is the 'surface current' (SC) model and is the one followed in this paper. Experiments carried out at energies from a few MeV to 28.5 GeV appear to be in agreement with this theory [35-38]. A recent review of the various models for SP radiation can be found in [39] where the authors also suggest another model, based on the similarities between SP and diffraction radiation.

The predictions of the EFIE and the SC models are not in good agreement with each other and a detailed comparison is still pending. The issue will probably not be resolved until more data become available. Nevertheless, it is true to say that the difference in predicted output is less than one order of magnitude and that this level of agreement is adequate for the purposes of this work. It is important to note that a common assumption made by both theories is that the grating has perfect conductivity in the sense that the electrons are so mobile that they can adjust immediately to the changes of the electric field, irrespective of its frequency.

## III. CALCULATION OF RADIATED ENERGY.

According to the SC theory of SP radiation, the energy per solid angle generated by the passage of a single electron over a grating of length *Z* and period *l* is given by the following semi-analytical expression [33, 34, 39], in CGS units:

$$\left(\frac{dI}{d\Omega}\right)_1 = 2\pi e^2 \frac{Z}{l^2} \frac{n^2\beta^3}{(1-\beta\cos\theta)^3} \exp\left[-\frac{2x_0}{\lambda_e}\right] R^2 \qquad (2)$$

where $\theta$ is the observation angle relative to the beam direction, $\beta$ is the relativistic factor, $x_0$ is the height of the electron above the grating surface and $\lambda_e$ is the evanescent wavelength, which is a measure of the coupling efficiency between the electron and the grating and is given by:

$$\lambda_e = \frac{\beta\gamma\lambda}{2\pi\sqrt{1+\beta^2\gamma^2\sin^2\theta\sin^2\phi}}$$

The dimensionless quantity $R^2$ is the grating efficiency factor, which is a complicated function of the blaze angle of the grating and is calculated numerically [35]. The energy radiated by a bunch containing $N_e$ electrons and having a time profile given by $T$(t) is given by:

$$\left(\frac{dI}{d\Omega}\right)_{N_e} \approx \left(\frac{dI}{d\Omega}\right)_1 N_e^2 \left|\int_{-\infty}^{\infty} Te^{-i\omega t}dt\right|^2 \qquad (3)$$

It should be noted that the expression (2) is valid if the width of the grating is assumed to be infinite, in which case the dependence on $x_0$ is explicit and appears as an exponential factor. Otherwise, it is necessary to resort to numerical integrations of modified Bessel functions $K_1$ over the finite width *w* of the grating.

The basic assumptions for the calculations that follow are:

- The grating material is perfectly conducting.
- The emitted radiation is observed at ‘infinity’, i.e. at a distance that is large compared with the length of the grating.
- The *x*, *y* and *t* charge distributions in the bunch are uncorrelated and are Gaussian in the transverse directions (*x*, *y*).

- Unless otherwise stated, the width of the grating (*y*-direction) is infinite. This approximation reduces the computational time but it does introduce an error in the case of highly relativistic beams, where the field lines of the electron will be spread out in the *y*-direction. A limited number of simulations with gratings of finite width have been carried out and are shown for comparison purposes.
- All calculations are for first order emission ($n$=1).

The basic set of beam parameters that have been used for the calculations presented in this section are listed in Table I.

Table I: Beam parameters

| | |
|---|---|
| Relativistic factor ($\gamma$) | 1000 |
| Electrons per bunch ($N_e$) | $10^9$ |
| Transverse size ($\sigma_x$, $\sigma_y$) | 0.02 mm |
| Bunch time profile | Symmetric Gaussian |
| Bunch length (FWHM) | 50 fs= 15 $\mu$m |

**1. Grating selection**.

The three criteria for the selection of the optimum grating periods were: (a) adequate radiated energy (b) the achievement of the widest possible wavelength coverage in order to aid the bunch shape reconstruction and (c) the ability to discriminate between different bunch shapes. The definition of 'adequate' will depend on the choice of detector(s). This point is discussed in some detail in section IV. For the case of a 50 fs bunch (FWHM), the three grating periodicities that were found to give a reasonable compromise between these requirements were $l$=15, 85 and 500 $\mu$m. The output from any grating will depend, amongst other factors, on the choice of blaze angle. A series of simulations indicated that the optimal blaze angles were $30^0$, $30^0$ and $35^0$, respectively. The expected angular distribution of SP radiation from these gratings is shown in Fig. 2, which shows the differential energy (Joules per steradian per cm of grating length) as a function of the observation angle ($\theta$). However, collection of SP radiation in the very forward (or backward) directions is rather difficult and, based on the experience of recent experiments [37, 38], the calculations have been truncated to observation angles ($\theta$) in the range 40-140$^0$ (shaded region in the figure). Fig. 3 shows, with solid lines, the same results as Fig. 2, but plotted against wavelength. The wavelength coverage achieved by this combination of gratings is 10-1000 $\mu$m, approximately, which is satisfactory. For comparison purposes, the results of the calculation for a grating of finite width (20 mm in this case) are also shown in Fig. 3 by symbols.  The difference between the two calculations is negligible for wavelengths lower than about 40 $\mu$m but it does rise to a factor of 2, approximately, in the longer wavelength region.

All the above calculations have been made on the assumption that the longitudinal (time) profile of the bunch is a simple, symmetric Gaussian shape. Fig. 4 shows the changes in the spectral distribution that will occur if the shape of the bunch is an asymmetric Gaussian, with an asymmetry factor $\epsilon$=2, which means that the trailing part of the bunch profile has a value of $\sigma$ that is twice as long as that of the leading edge; the FWHM of the bunch is unchanged, at 50 fs. The solid lines represent the yield from the symmetric shape and the dashed ones from the asymmetric bunch. It is evident that if bunch shape discrimination is the main consideration, then the information provided by the short period grating is important. This can be understood in simple, qualitative terms as follows: the differences in shape will be evident at the onset of the coherent emission, when the wavelength is a bit shorter than the length of the bunch. Conversely, when the wavelength becomes significantly longer than the bunch length, the bunch is seen as a 'lump' of charge and all information about its shape is lost. This emphasises the importance of obtaining information both from the long and the short period gratings.

**2. Beam position and size**.

Fig. 5 shows the expected variation of the radiated energy as a function of the height of the beam ($x_0$) above the grating surface. The curves have been normalised to the $x_0$=1.0 mm value. The simulation has been carried out for an observation angle of $\theta$=90$^0$ where the wavelength is equal to the period of the grating. As expected, the energy decreases almost exponentially as the beam-grating separation increases. This is more pronounced for shorter wavelengths, where good coupling with the grating requires the beam to be close to the grating surface. The assumed beam-grating separation of $x_0$=1.0 mm is a reasonable compromise between having adequate signal while avoiding the region of very rapid variation with beam height.

The effect of varying the transverse size of the beam is shown in Fig. 6 where it has been assumed that a round beam, with Gaussian profiles in the *x* and *y* dimensions, is travelling over a grating of $l$=15 $\mu$m, at a height of 1.0 mm above it. The observation angle is $120^0$, hence the wavelength is equal to 22.5 $\mu$m. The effect is minimal, up to $\sigma_x=\sigma_y$=0.3 mm, approximately. Similar results have been obtained from the other gratings. The case of 'flat' beams ($\sigma_x \ll \sigma_y$) has also been considered and it was concluded, again, that there is minimal change to the output for $\sigma_y$<0.3 mm, approximately.

### 2. **Beam energy**.

It is also interesting to consider the effect of changing the energy of the electron beam. We have considered energies ranging from $\gamma$=1000 (E≅500 MeV) down to $\gamma$=20 (E≅10 MeV). The results of the simulations are shown in Fig. 7, plotted as $J.sr^{-1}.cm^{-1}$ vs. wavelength. The grating period was 85 $\mu$m and the assumed beam height above the grating was 1.0 mm. A change in the value of $\gamma$ from 1000 down to 500 does not have a significant effect on the level of SP radiation. However, further reductions in $\gamma$ cause a drastic drop in the SP energy. This is particularly noticeable in going down from $\gamma$=100 to $\gamma$=20. The question of detectability of the radiation at low beam energies is discussed in the following sections.

## IV. EXPERIMENTAL ISSUES.

The three main topics for consideration on the experimental side are detectors, filters and the availability of suitable gratings. The discussion is based on recent experience at SLAC [38], where an array of 11 pyroelectric (PE) detectors was used to detect SP radiation, but at wavelengths in the region between 0.5 and 2.6 mm, approximately. It should be noted that the arrangement envisaged in this paper covers a wide part of the spectrum, from the mid-infrared up to the sub-millimetre region and this presents some additional complications. These issues are considered, in turn, in the following sections.

## 1. Detectors.

The pyroelectric detector is certainly capable of covering the wavelength range considered here, but the question is whether one could do better with another type of detector. A partial list of the infrared detectors, currently available, is shown in Fig. 8, together with their noise equivalent power (NEP), wavelength range and their respective operating temperatures. This is a very simplified picture that does not include any variations in detectivity with wavelength and it also ignores the fact that in certain cases the NEP of the detector itself may not be the dominant factor. This was the case with the detector/electronics system of [38], where the main source of NEP was the electronics. Liquid helium cooled detectors have been excluded because of the cost and complexity of a cryogenic system but they are always an option. The list does include detectors operating at 77K, which do not present such problems. The Golay cell has a very slow response time, is rather bulky and expensive and can also be excluded from consideration. This leaves the pyroelectric as the only detector covering the whole range of wavelengths. One possible alternative, covering wavelengths up to about 20 $\mu$m, is the HgCdTe photodiode, which operates at liquid nitrogen temperature and with a NEP about two orders of magnitude better than that of the pyroelectric one. With carefully designed electronics, it is reasonable to expect an improvement in the NEP of about 3 orders of magnitude over that of the PE detector. In view of the previously mentioned importance of short wavelength information for bunch shape discrimination, this type of detector could be very useful.

In order to convert the calculated differential energy of Fig. 3 into energy incident on the detector, we assume the same parameters as [38], namely a grating that is 4 cm long and an optical system that accepts a solid angle of $6.5 \times 10^{-3}$ sr. This is shown in Fig. 9. The solid line marked PE represents the experimentally established detection limit, for this type of detector. It would be possible to improve on this by paying particular attention to the electronics but, assuming for the moment that this is indeed the limit of detection for the PE detector, then it is obvious that the measurements suggested in this paper can be carried out by an array of PE detectors. If the bunch charge were to drop to $10^8$ electrons per bunch, then the use of HgCdTe detectors would become necessary for the shorter wavelengths.

## 2. **Filters and gratings**.

The use of filters in order to discriminate against background radiation is absolutely essential for any SP experiment. Discrimination against long wavelength background is particularly important, but we do not foresee any difficulties in obtaining suitable band pass filters for the wavelength region considered here. The longer period grating considered here (0.5 mm) can be manufactured in any good workshop. The shorter period ones can be purchased from specialist manufacturers. In order to ensure 'perfect' electrical conductivity, the gratings should be made on a solid metal substrate; aluminium or copper would be suitable for this purpose.

# V. SUMMARY AND DISCUSSION

In the present paper we have considered the theoretical and practical issues associated with the use of coherent Smith-Purcell radiation for the determination of the time profile of very short (a few tens of femtoseconds) electron bunches. Amongst the various radiative processes that could be used for this purpose, SP radiation is particularly attractive because it offers wide spectral coverage through the simultaneous use of multiple gratings, with different periodicities. This, in turn, facilitates the bunch profile reconstruction by Kramers-Kronig analysis. Therefore, coherent SP radiation is a relatively inexpensive alternative or complementary technique to the transverse deflecting cavities or to Electro-Optic sampling.

The use of a radiative process, such as SP radiation, for the determination of the time profile of a bunch requires a theoretical framework for the calculation of the expected spectral distribution of the radiation. The calculations presented here have been carried out on the basis of the 'surface current' model for the emission of SP radiation, which is in broad agreement with the latest experimental results in the far infrared and sub-millimetre wavelength region. One of the main assumptions of this formulation of the problem (and, indeed, of the alternative EFIE formulation) is that the grating surface is perfectly conducting. Although this is certainly a valid assumption for the far infrared, it has to be re-examined for bunch lengths of the order of 50 fs when the emitted radiation reaches down to the near infrared. The shortest wavelength considered here ($\lambda$=10 $\mu$m) corresponds to a frequency $\omega$= 1.9x$10^{14}$ $s^{-1}$. The published data indicate rather modest decrease in conductivity up to this frequency but beyond that, the decrease in conductivity is more rapid [40]. The optical properties of aluminium can be described in terms of the Drude model and the scattering and plasma frequencies for this material lie in the region of $10^{14}$ $s^{-1}$ and $10^{16}$ $s^{-1}$, respectively [40]. Therefore, for aluminium and for frequencies lower than $10^{14}$ $s^{-1}$, approximately, the conductivity of the metal is essentially independent of frequency. These comments are in line with the observation that the finite conductivity of the metal is not an issue for wavelengths down to about 4 $\mu$m [41]. It should be noted that the scattering (or 'damping') frequency depends on lattice imperfections and impurities [42] and, thus, on the condition of the grating surface. Hence, gratings made out of solid metal are, in principle, preferable. The conclusion from this brief discussion is that the existing version of the surface current model of Smith-Purcell radiation will give reasonably accurate estimates of radiated energies for the spectral range considered in the present study and it may even be applicable to slightly shorter bunches of the order of 10 fs. For even shorter bunch lengths, the present theories for the emission process would have to be modified in order to account for the decreasing surface conductivity of the metal.

In conclusion, the work reported here indicates that there are no obstacles, either theoretical or experimental, to the use of coherent SP radiation for the single-shot measurement of the bunch length and (time) profile of 50 fs long bunches. Using realistic beam parameters and assuming an experimental arrangement that can accommodate three gratings with periods of 15, 85 and 500 $\mu$m, it was shown that an array of inexpensive, room temperature pyroelectric detectors is capable of measuring the SP radiation over the range of 10-1000 $\mu$m, provided the bunch charge is of the order of $10^9$ electrons/bunch. Measurements over this range should be adequate for the reconstruction of the bunch profile. Greater sensitivity could be achieved through the use of HgCdTe detectors, operating at liquid nitrogen temperature. These detectors, however, have a limited range and would only be useful for wavelengths up to about 20 $\mu$m.

**Acknowledgements**

The authors are grateful to Prof. M.F. Kimmitt and to Drs. V. Blackmore and A. Kesar for helpful discussions and comments. The financial support of the Oxford University Fell Fund is gratefully acknowledged. One of us (E.M) would also like to acknowledge the support of Hertford College, Oxford.

## REFERENCES

1. W. P. Leemans, B. Nagler, A. J. Gonsalves, Cs. Tóth, K. Nakamura, C. G. R. Geddes, E. Esarey, C. B. Schroeder and S. M. Hooker, Nature Physics **2**, 696, (2006).
2. J. Faure, Y. Glinec, A. Pukhov, S. Kiselev, S. Gordienko, E. Lefebvre, J.-P. Rousseau, F. Burgy & V. Malka, Nature **431,** 541, (2004).
3. O.H. Altenmueller, R.R. Larsen and G.A. Loew, Rev. Sci. Instrum., **35**, 238, (1964).
4. L.Fröhlich, O.Grimm et al, **SLAC-PUB-11387.**
5. M. Hüning, A. Bolzmann, H. Schlarb et al, **SLAC-PUB-11482.**
6. G. Berden, W.A. Gillespie, S.P. Jamison, E.A. Knabbe, A.M. MacLeod, A.F.G. van der Meer, P.J. Phillips, H. Schlarb. B. Schmidt, P.Schmüser and B. Steffen, Phys. Rev. Lett. **99**, 164801 (2007).
7. B. Steffen , V. Arsov, G. Berden, W.A. Gillespie, S.P. Jamison, A.M. MacLeod, A.F.G. van der Meer, P.J. Phillips, H. Schlarb. B. Schmidt and P.Schmüser, Phys. Rev. Special Topics-Accelerators & Beams **12,** 032802 (2009).
8. S. J. Smith and E. M. Purcell, Phys. Rev. **92**, 1069, (1953).
9. J.C. McDaniel, D.B. Chang, J.E. Drummond and W.W. Salisbury, Applied Optics, **28**, 4924, (1989).
10. G. Doucas, J.H. Mulvey, M. Omori, J. Walsh and M.F. Kimmitt, Phys. Rev. Lett., **69**, 1761, (1992).
11. K.J. Woods, J.E. Walsh, R.E. Stoner, H.G. Kirk and R.C. Fernow, Phys. Rev. Lett., **74**, 3808, (1995).
12. J.H. Brownell, J. Walsh, H.G. Kirk, R.C. Fernow and S.H. Robertson, Nucl. Instrum. & Methods in Phys. Research A, **393**, 323, (1997).
13. Y. Shibata, S. Hasebe, K. Ishi, S. Ono, M. Ikezawa, T. Nakazato, M. Oyamada, S. Urasawa, T. Takahashi, T. Matsuyama, K. Kobayashi and Y. Fujita, Phys. Rev. E, **57**, 1061, (1998).

14. J. Urata, M. Goldstein, M.F. Kimmitt, A. Naumov, C. Platt and J.E. Walsh, Phys. Rev. Lett., **80**, 516, (1998).
15. H.L. Andrews, C.H. Boulware, C.A. Brau and J.D. Jarvis, Phys. Rev. Special Topics-Accel. & Beams **8**, 110702, (2005).
16. J.T. Donohue and J. Gardelle, Phys. Rev. Special Topics-Accel. & Beams **8**, 060702, (2005).
17. S.E. Korbly, A.S. Kesar J.R. Sirigiri and R.J. Temkin, Phys. Rev. Lett. **94**, 054803, (2005).
18. A. S. Kesar, R.A. Marsh and R.J. Temkin, Phys. Rev. Special Topics- Accelerators & Beams **9,** 022801 (2006).
19. H.L. Andrews, C.A. Brau, J.D. Jarvis, C.F. Guertin, A. O'Donnell, B. Durant, T.H. Lowell and M.R. Mross, Phys. Rev. Special Topics-Accel. & Beams **12**, 080703, (2009).
20. R. Lai and A.J. Sievers, Phys. Rev. E, **50**, R3342, (1994).
21. R. Lai, U. Happek and A.J. Sievers, Phys. Rev. E, **50**, R4294, (1994).
22. R. Lai and A.J. Sievers, Nucl. Instrum. & Methods in Phys. Research A, **397**, 221 (1997).
23. O. Grimm and P. Schmüser, TESLA FEL Report 2006-03.
24. V. Blackmore, D.Phil. Thesis, University of Oxford, Michaelmas Term 2008, unpublished.
25. G. Toraldo di Francia, Nuovo Cimento **16**, 61, (1960).
26. P.M. van den Berg, J. Opt. Soc. of America, **63**, 689, (1973).
27. P.M. van den Berg, J. Opt. Soc. of America, **63**, 1588, (1973).
28. O. Haeberlé, P. Rullhusen, J.-M. Salomé and N. Maene, Phys. Rev. E **49**, 3340, (1994).
29. G. Kube, Nucl. Instrum. & Methods in Phys. Research B, **227**, 180, (2005).
30. A. S. Kesar, M. Hess, S. Korbly and R. Temkin, Phys. Rev. E **71**, 016501, (2005).
31. A.S. Kesar, Phys. Rev. Special Topics- Accelerators & Beams, **8**, 072801, (2005).
32. J. H. Brownell, J. E. Walsh and G. Doucas, Phys. Rev. E **57**, 1075, (1998).
33. S.R. Trotz, J.H. Brownell, J.E. Walsh and G. Doucas, Phys. Rev. E **61**, 7057, (2000).
34. J.H. Brownell and G. Doucas, Phys. Rev. Special Topics- Accelerators & Beams **8,** 091301 (2005).
35. G. Doucas, J.H. Mulvey, M. Omori, J. Walsh and M.F. Kimmitt, Phys. Rev. Lett., **69**, 1761, (1992).
36. G. Doucas, M.F. Kimmitt, A. Doria, G.P. Gallerano, E. Giovenale, G. Messina H.L. Andrews and J.H. Brownell, Phys. Rev. Special Topics- Accelerators & Beams **5,** 072802 (2002).

37. G. Doucas, V. Blackmore, B. Ottewell, C. Perry, P.G. Huggard, E. Castro-Camus, M.B. Johnston, J. Lloyd-Hughes, M.F. Kimmitt, B. Redlich and A. van der Meer, Phys. Rev. Special Topics- Accelerators & Beams **9,** 092801, (2006).
38. V. Blackmore, G. Doucas, C. Perry, B. Ottewell, M.F. Kimmitt, M. Woods, S. Molloy and R. Arnold, Phys. Rev. Special Topics- Accelerators & Beams **12,** 032803, (2009).
39. D.V. Karlovets and A.P. Potylitsyn, Phys. Rev. Special Topics- Accelerators & Beams **9,** 080701, (2006).
40. M.A. Ordal, R.J. Bell, R.W. Alexander Jr., L.A. Newquist and M.R. Querry, Applied Optics **27**, 1203, (1988).
41. C. Palmer, 'Diffraction Grating Handbook', Newport Corporation, 6th Edition, 2005, Ch.9, http://gratings.newport.com/library/handbook/toc.asp.
42. J.D. Jackson, 'Classical Electrodynamics', John Wiley & Sons, 2nd Edition, 1975, Ch.7.

**Figure Captions**

Fig. 1. Schematic of the Smith-Purcell radiative process and definition of the coordinate system.

Fig. 2. Angular distribution of the radiated energy per solid angle and per unit of grating length, for three different gratings. The shaded area is the one that is assumed to be experimentally accessible.

Fig. 3. The differential energy of Fig.2 plotted against wavelength, for a grating of infinite width (solid lines) and for a 20mm wide grating (symbols). The wavelength range corresponds to the shaded area of Fig.2.

Fig. 4. Calculated spectral distributions from two different bunch shapes, both 50fs (FWHM) long. The solid lines are for a symmetric ($\epsilon$=1) Gaussian bunch and the dashed ones for a bunch whose trailing side has a sigma that is twice as long as that of the leading edge ($\epsilon$=2).

Fig. 5. Calculated effect of beam position variation on the radiated energy (see text for details). The output has been normalised to the value expected at $x_0$=1.0mm.

Fig. 6. The effect of beam size variations in the transverse dimensions (*x* and *y*). See text for details.

Fig. 7. The effect of beam energy variation on the expected output from a grating having a period of 85$\mu$m. The assumed beam height above the grating is 1.0mm.

Fig. 8. A partial list of detectors covering the wavelength range 0.1-1000$\mu$m and their corresponding noise equivalent power (NEP). See text for details.

Fig. 9. Demonstration of the capability of pyroelectric (PE) detectors to respond to the energy levels expected in this study. The solid lines correspond to a bunch charge of $10^9$ electrons and the dashed ones to a charge of $10^8$. See text for details.

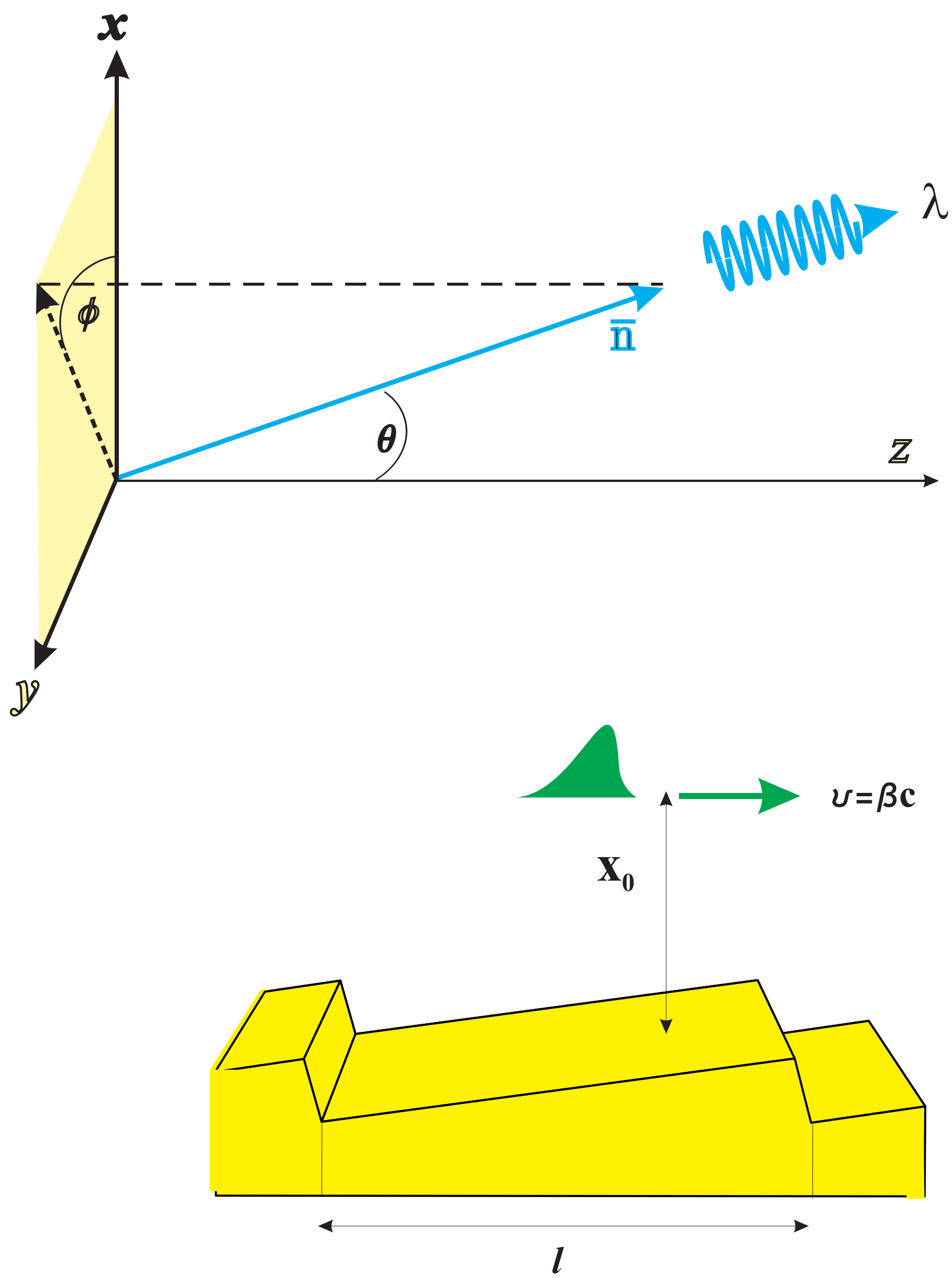

Fig.1

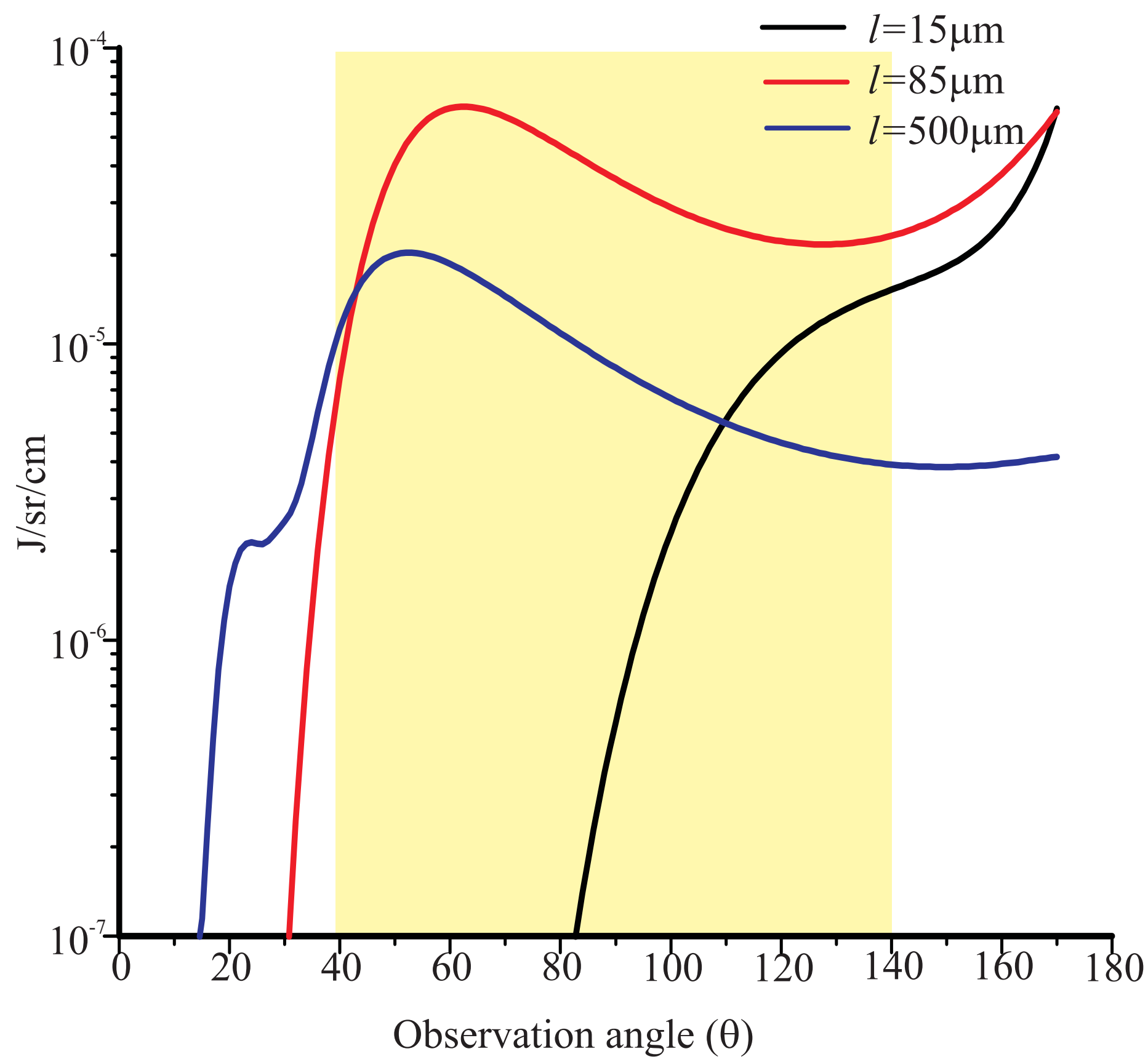

l=15μm
l=85μm
l=500μm
10-4
10-5
10-6
10-7
J/sr/cm
0
20
40
60
80
100
120
140
160
180
Observation angle (θ)

Fig. 2

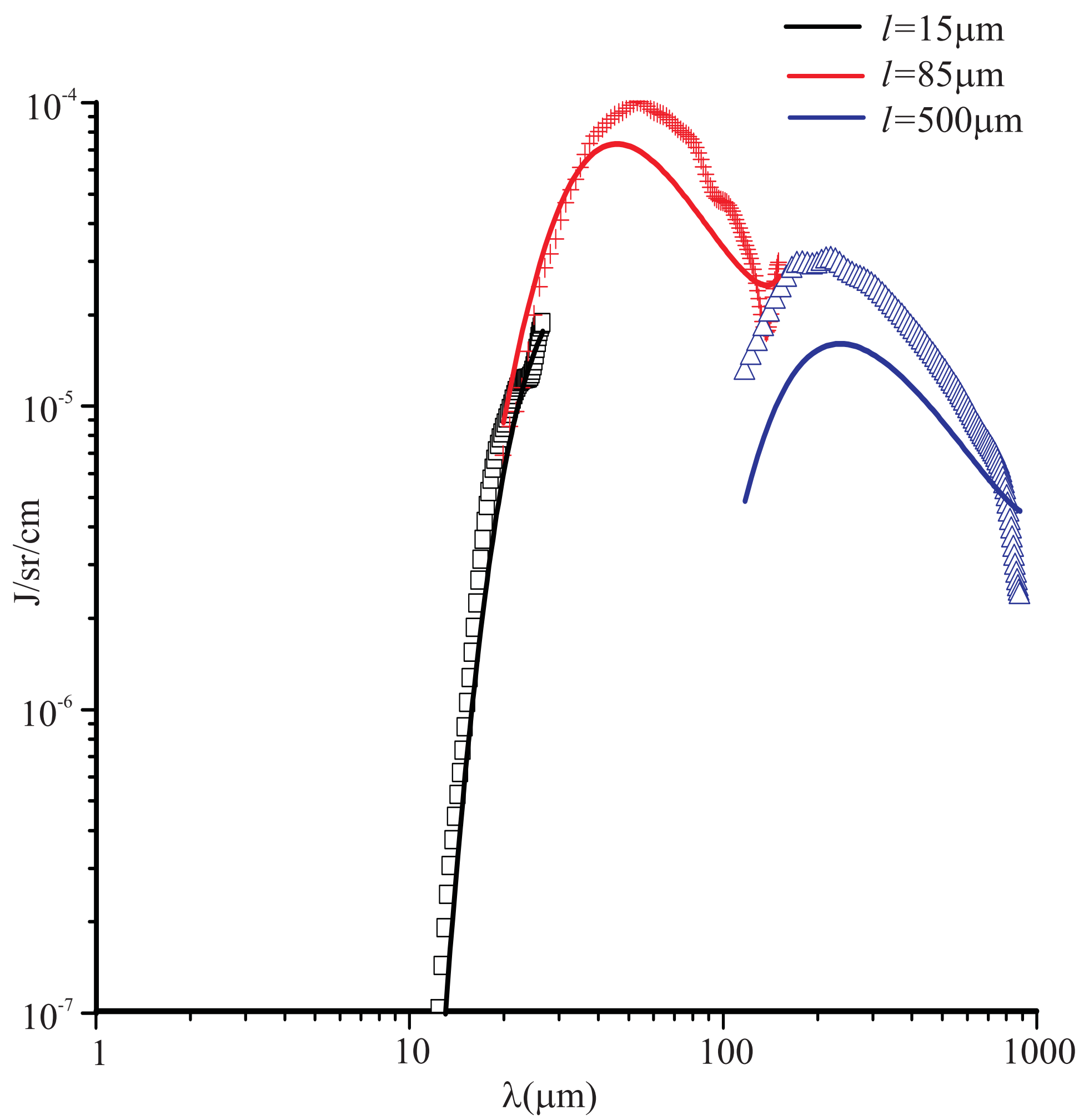

Fig. 3

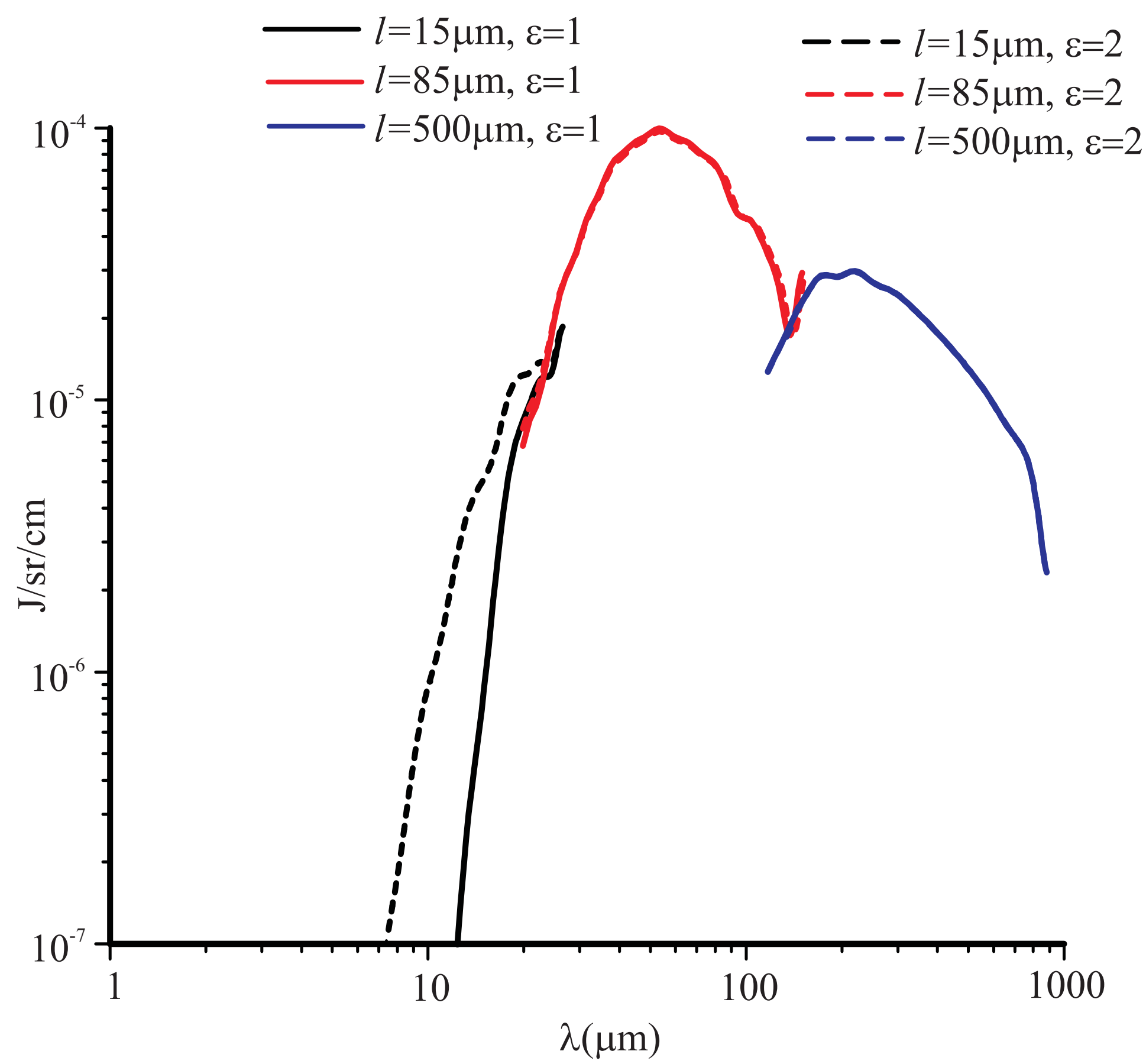

l=15μm, ε=1
l=85μm, ε=1
l=500μm, ε=1
l=15μm, ε=2
l=85μm, ε=2
l=500μm, ε=2
10^-4
10^-5
10^-6
10^-7
J/sr/cm
1
10
100
1000
λ(μm)

Fig. 4

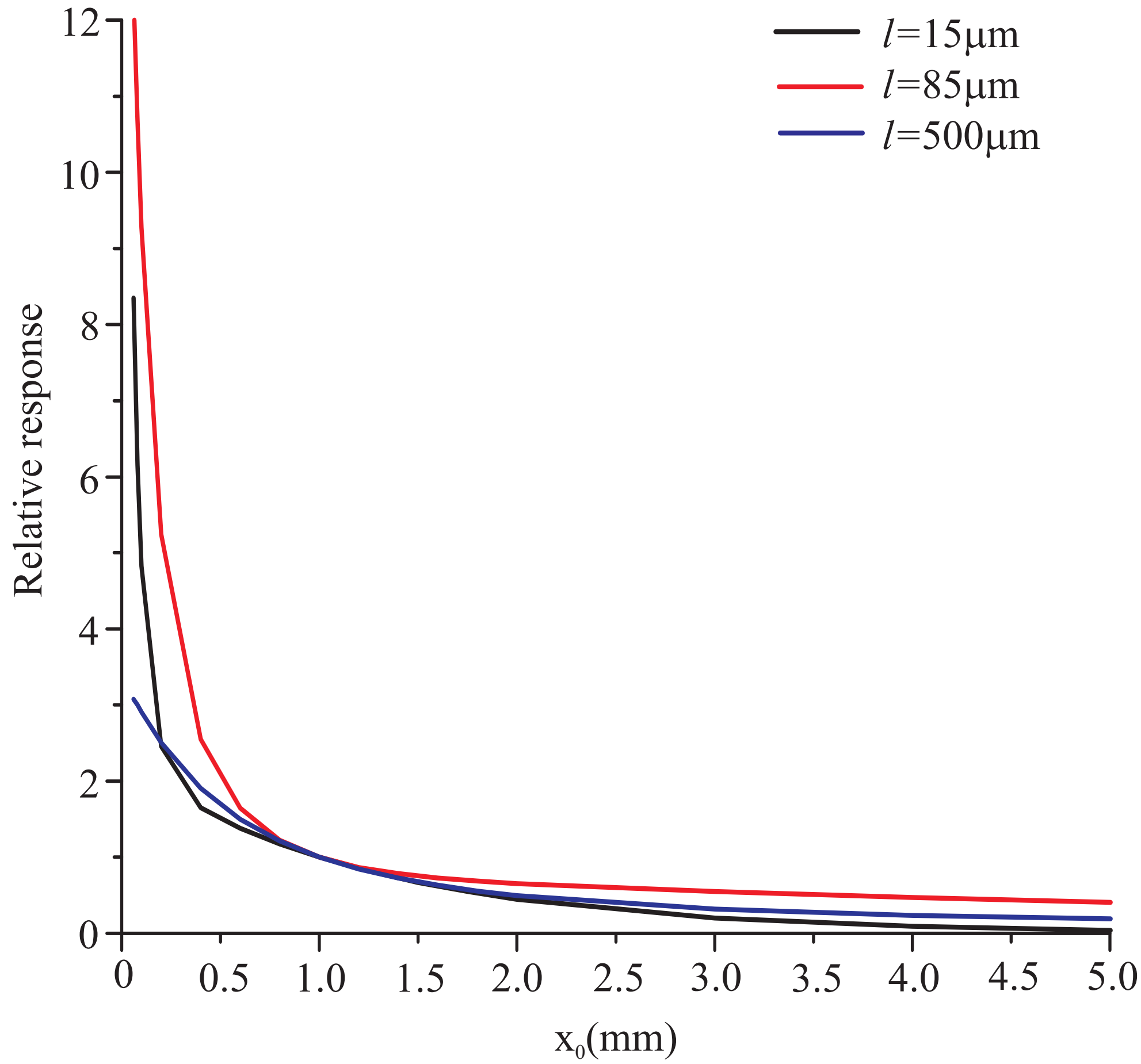

l=15μm
l=85μm
l=500μm
Relative response
0
2
4
6
8
10
12
0
0.5
1.0
1.5
2.0
2.5
3.0
3.5
4.0
4.5
5.0
x0(mm)

Fig. 5

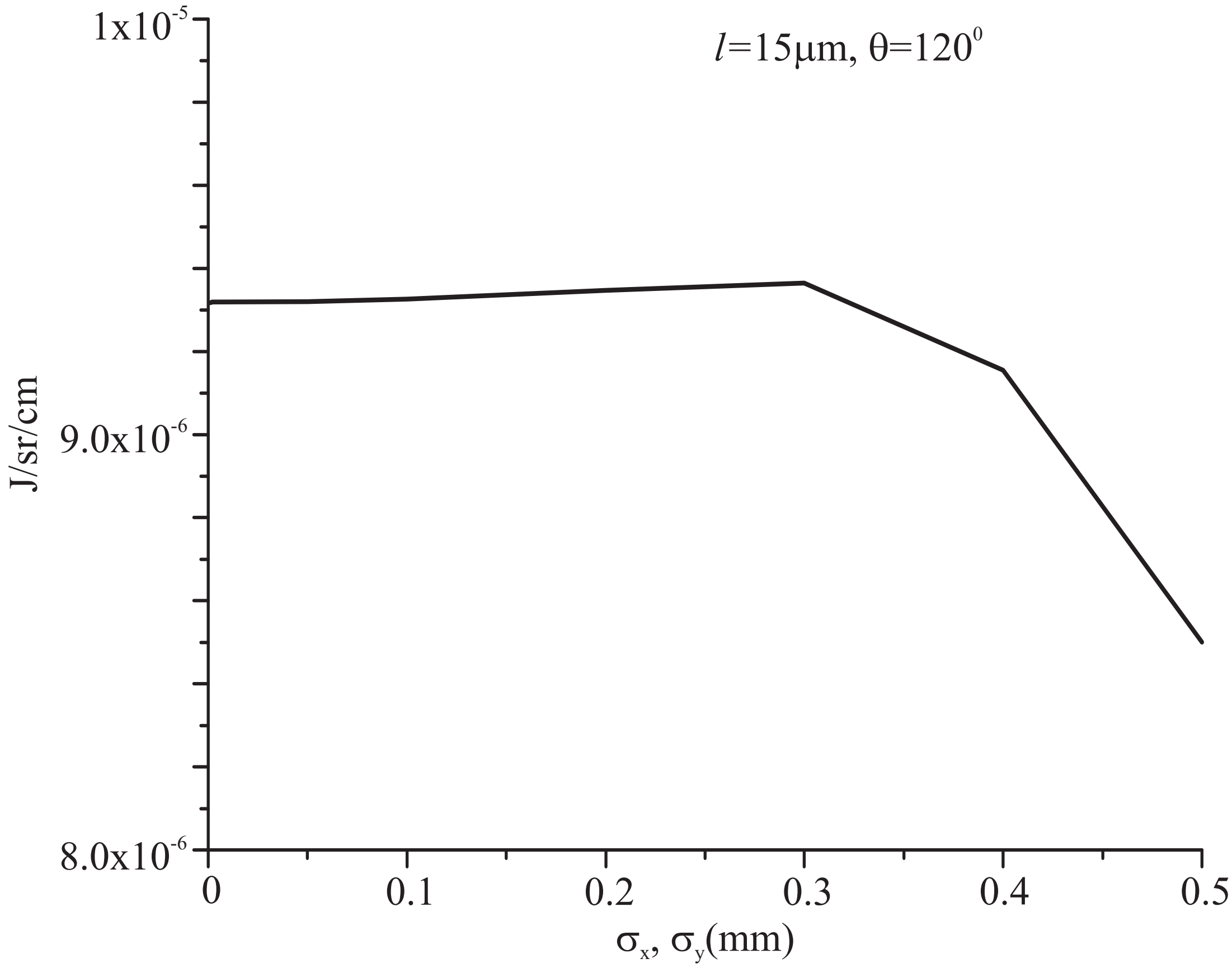

Fig. 6

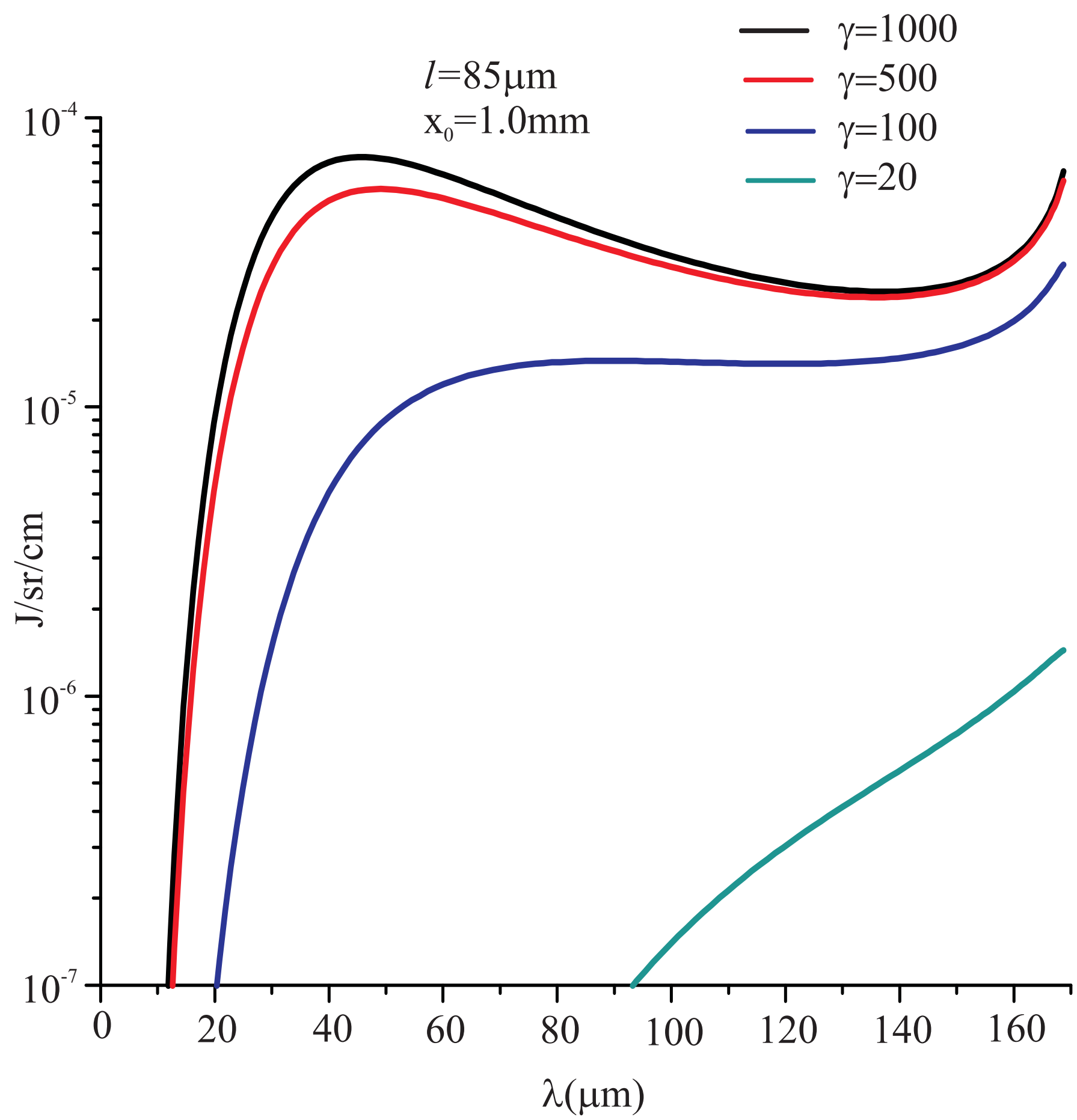

γ=1000
γ=500
γ=100
γ=20
*l*=85μm
x0=1.0mm
10-4
10-5
10-6
10-7
J/sr/cm
0
20
40
60
80
100
120
140
160
λ(μm)

Fig. 7

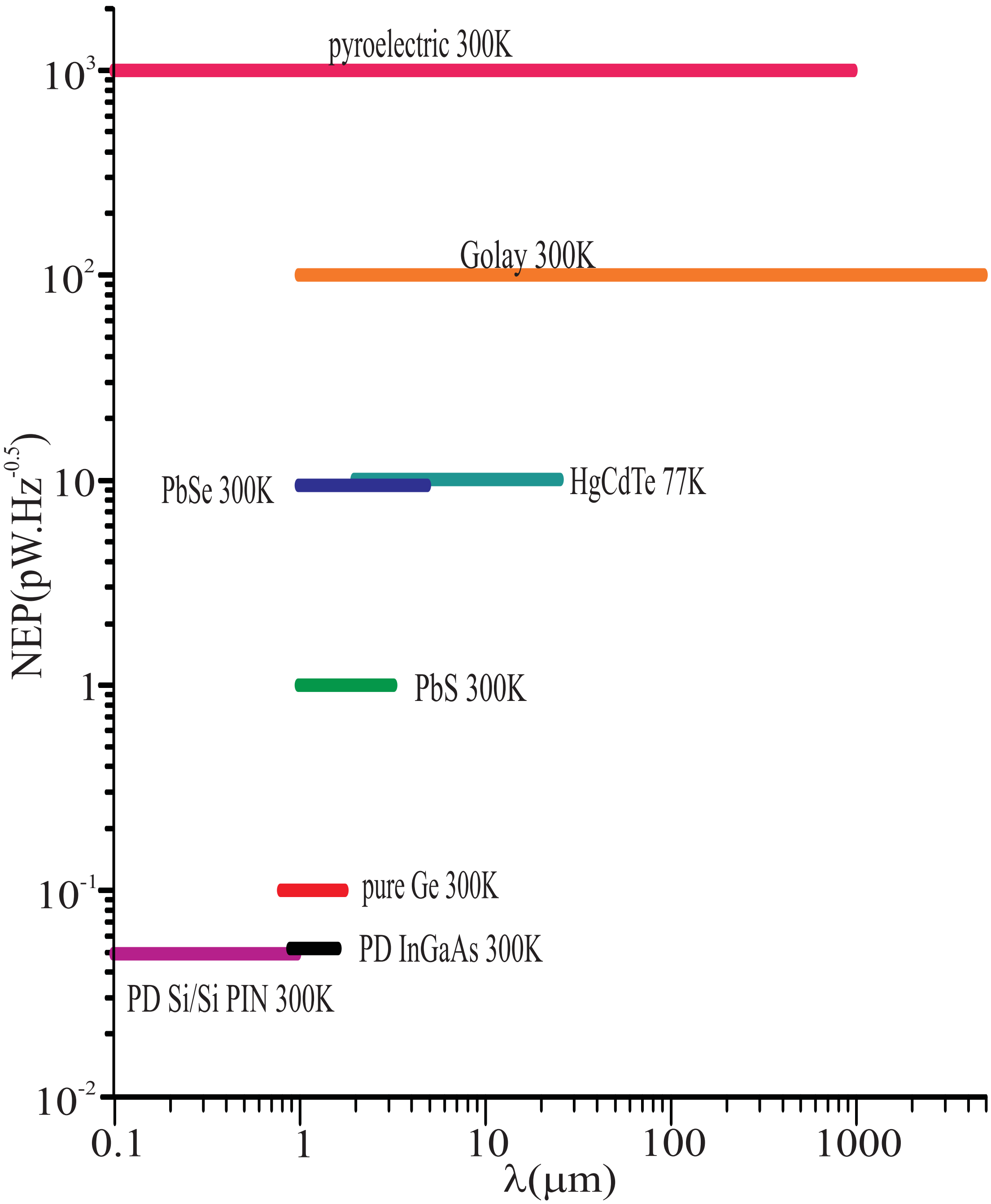
pyroelectric 300K
Golay 300K
PbSe 300K
HgCdTe 77K
PbS 300K
pure Ge 300K
PD InGaAs 300K
PD Si/Si PIN 300K
NEP(pW.Hz^-0.5)
10^3
10^2
10
1
10^-1
10^-2
0.1
1
10
100
1000
λ(μm)

Fig. 8

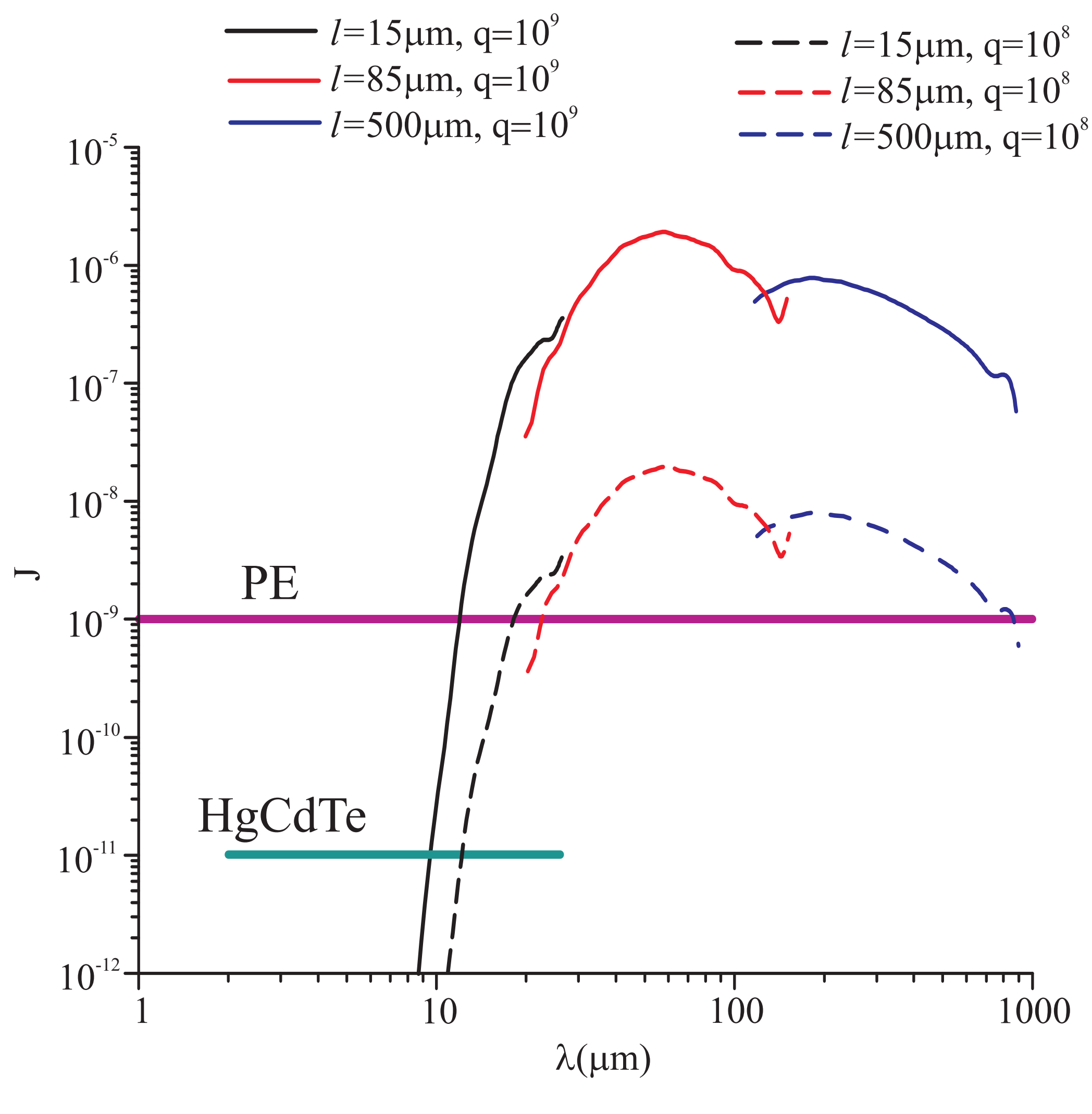

l=15µm, q=10^9
l=85µm, q=10^9
l=500µm, q=10^9
l=15µm, q=10^8
l=85µm, q=10^8
l=500µm, q=10^8
10^-5
10^-6
10^-7
10^-8
10^-9
10^-10
10^-11
10^-12
J
PE
HgCdTe
1
10
100
1000
λ(µm)

Fig. 9